\documentclass[aps,prb,superscriptaddress,twocolumn]{revtex4}
\usepackage{amssymb}
\usepackage{graphicx}
\usepackage{hyperref}

\begin{document}
\title{Single crystal growth and physical properties of the layered arsenide BaRh$_2$As$_2$}
\author{Yogesh Singh}
\affiliation{Ames Laboratory and Department of Physics and Astronomy, Iowa State University, Ames, IA 50011}
\author{Y. Lee}
\affiliation{Ames Laboratory and Department of Physics and Astronomy, Iowa State University, Ames, IA 50011}
\author{S. Nandi}
\affiliation{Ames Laboratory and Department of Physics and Astronomy, Iowa State University, Ames, IA 50011}
\author{A. Kreyssig}
\affiliation{Ames Laboratory and Department of Physics and Astronomy, Iowa State University, Ames, IA 50011}
\author{A. Ellern}
\affiliation{Department of Chemistry, Iowa State University, Ames, IA 50011}
\author{S. Das}
\affiliation{Ames Laboratory and Department of Physics and Astronomy, Iowa State University, Ames, IA 50011}
\author{R. Nath}
\affiliation{Ames Laboratory and Department of Physics and Astronomy, Iowa State University, Ames, IA 50011}
\author{B. N. Harmon}
\affiliation{Ames Laboratory and Department of Physics and Astronomy, Iowa State University, Ames, IA 50011}
\author{A. I. Goldman}
\affiliation{Ames Laboratory and Department of Physics and Astronomy, Iowa State University, Ames, IA 50011}
\author{D. C. Johnston}
\affiliation{Ames Laboratory and Department of Physics and Astronomy, Iowa State University, Ames, IA 50011}
\date{\today}

\begin{abstract}
Single crystals of BaRh$_2$As$_2$ have been synthesized from a Pb flux. We present the room temperature crystal structure, single crystal x-ray diffraction measurements as a function of temperature $T$, anisotropic magnetic susceptibility $\chi$ versus $T$, electrical resistivity in the $ab$-plane $\rho$ versus $T$, Hall coefficient versus $T$ and magnetic field $H$, and heat capacity $C$ versus $T$ measurements on the crystals. 
The single crystal structure determination confirms that BaRh$_2$As$_2$ forms in the tetragonal ThCr$_2$Si$_2$ type structure (space group \emph{I4/mmm}) with lattice parameters $a$~=~$b$~=~4.0564(6)~\AA~ and $c$~=~12.797(4)~\AA.  Band structure calculations show that BaRh$_2$As$_2$ should be metallic with a small density of states at the Fermi energy $N(E_{\rm F})$~=~3.49~states/eV~f.u. (where f.u.$\equiv$ formula unit) for both spin directions.  $\rho(T)$ data in the $ab$-plane confirm that the material is indeed metallic with a residual resistivity $\rho(2~{\rm K})$~=~29~$\mu \Omega$~cm, and with a residual resistivity ratio $\rho(310~{\rm K})$/$\rho(2~{\rm K})$~=~5.3.  The observed $\chi(T)$ is small ($\sim 10^{-5}$~cm$^3$/mol) and weakly anisotropic with $\chi_{\rm ab}$/$\chi_{\rm c} \approx$~2.  The $C(T)$ data indicate a small density of states at the Fermi energy with the low temperature Sommerfeld coefficient $\gamma$~=~4.7(9)~mJ/mol~K$^2$.  There are no indications of superconductivity, spin density wave, or structural transitions between 2~K and 300~K\@.  We compare the calculated density of states versus energy of BaRh$_2$As$_2$ with that of BaFe$_2$As$_2$.

\end{abstract}
\pacs{65.40.Ba, 61.05.cp, 71.20.-b,  74.70.Ad,}

\maketitle

\section{Introduction}
\label{sec:INTRO}
The recent discovery of high temperature superconductivity in LaFeAsO$_{1-x}$F$_x$ with  $T_{\rm c}$~=~26~K (Ref.~\onlinecite{kamihara2008}) has generated a lot of activity in the search for related superconductors.  Even higher $T_{\rm c}$s of 35~K to 55~K have been achieved when La is replaced by other rare-earth elements $R$~=~Ce, Pr, Nd, Sm, Gd, Tb, and Dy.\cite{XHChen2008,GFChen2008,Ren2008,Ren22008,Yang2008,Bos2008,Cheng2008}   These materials form in the tetragonal ZrCuSiAs type structure (\emph{P4/nmm}).\cite{Quebe2000}  The structure is made up of FeAs layers alternated by LaO layers stacked along the $c$-axis.  The undoped materials $R$FeAsO show a spin density wave (SDW) transition and a coupled structural transition at high temperatures ($T \sim$150~K).\cite{Dong2008,GFChen2008, Klauss2008}  On doping with F, the SDW and the structural transitions are suppressed and superconductivity is observed.\cite{XHChen2008,GFChen2008,Ren2008,Ren22008,Yang2008,Cheng2008,Dong2008,Klauss2008,Giovannetti2008}  

Recently a different family of compounds $A$Fe$_2$As$_2$ ($A$~=~Ba, Sr, Ca, and Eu) was discovered which crystallize in the tetragonal ThCr$_2$Si$_2$ type structure and has similar FeAs layers seperated by $A$ layers stacked along the $c$-axis.  Like the $R$FeAsO materials, these materials also show SDW and structural transitions at high temperatures..\cite{Rotter2008,Krellner2008,Ni2008, Yan2008,Ni2008a,Ronning2008,Goldman2008,Tagel2008,Ren32008,Jeevan2008}   When the $A$ atoms are partially replaced by K, Na, or Cs, the SDW and structural transitions are suppressed and superconductivity is observed.\cite{2Rotter2008,2GFChen2008,2Jeevan2008,Sasmal2008}  The common feature in all the materials mentioned above is the FeAs layer in their crystal structure. 
It is of interest to look for other materials with related structures and investigate their physical properties to see if these can be potential parent compounds for new high temperature superconductors.  

\begin{figure}[t]
\includegraphics[width=3.in]{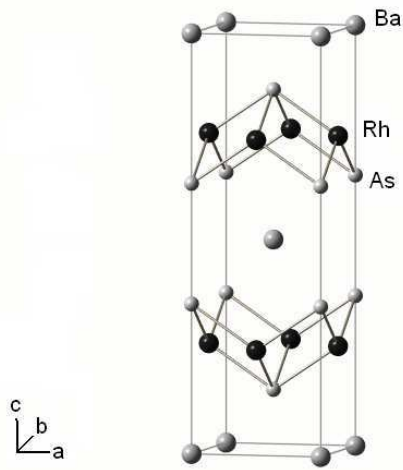}
\caption{The crystal structure of BaRh$_2$As$_2$.  Layers of Ba and layers of RhAs alternate along the $c$-axis.  Within the RhAs layers the As atoms lie outside the plane formed by the square-lattice Rh atoms. 
\label{Figstructure}}
\end{figure}

The series of compounds $A$Rh$_2$As$_2$ ($A$~=~Ba, Sr, and Eu) were previously synthesized in single crystal form and their crystal structure was reported.\cite{Hellmann2007}  The Ba and Eu compounds form in the ThCr$_2$Si$_2$ type structure while the Sr compound forms in a different structure.  The crystal structure of BaRh$_2$As$_2$ is shown in Fig.~\ref{Figstructure}.  The structure is built up of RhAs layers alternated by Ba layers stacked along the crystallographic $c$-axis.  Within the RhAs layers the As atoms lie outside the plane formed by the Rh atoms.  To the best of our knowledge the physical properties of these $A$Rh$_2$As$_2$ compounds have not been investigated before.

Herein we report on the crystal growth, single crystal structure, resistivity $\rho$ in the $ab$-plane versus temperature $T$, Hall effect versus $T$ and magnetic field $H$, magnetic susceptibility $\chi$ versus $T$, and heat capacity $C$ versus $T$ measurements of BaRh$_2$As$_2$.  Our experimental results are compared with predictions of band structure calculations.

\begin{figure}[t]
\includegraphics[width=3in]{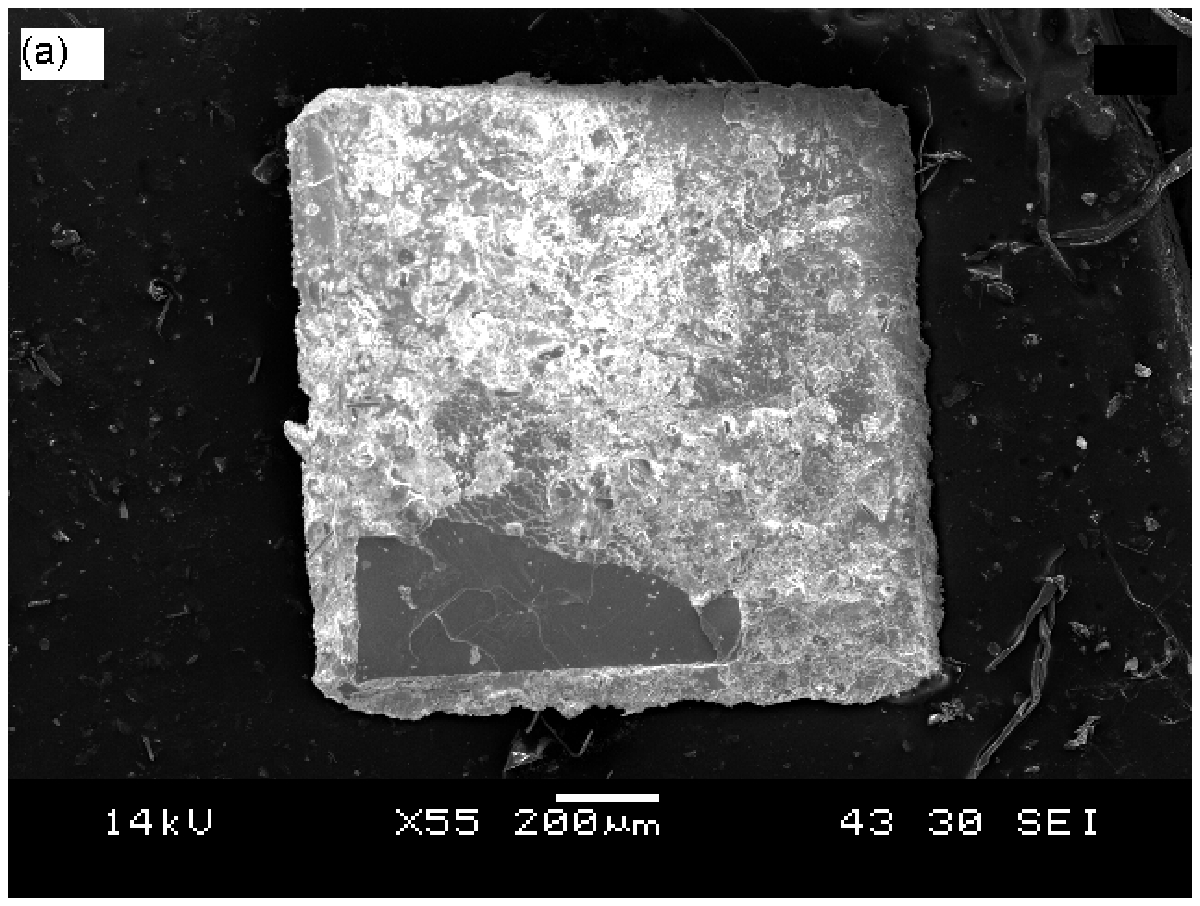}
\includegraphics[width=3in]{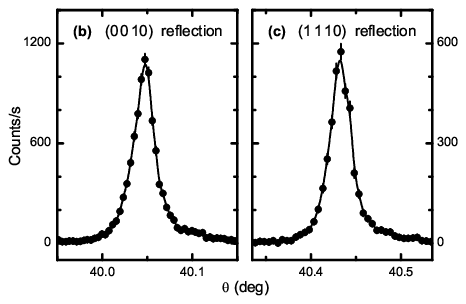}
\caption{(a) A scanning electron microscope (SEM) image of a typical as grown crystal of BaRh$_2$As$_2$.  The white mottled area on the surface is a thin layer of Pb from the flux.  (b) Rocking scan through the (0 0 10) reflection, and (c) rocking scan through the (1 1 10) reflection of a BaRh$_2$As$_2$ single crystal.   
\label{Fig-Xtal}}
\end{figure}

\section{Experimental DETAILS}
\label{sec:EXPT}
Single crystals of BaRh$_2$As$_2$ were grown out of Pb flux.\cite{Hellmann2007}  The elements were taken in the ratio Ba:Rh:As:Pb~=~ 1.1~:~2~:~2.1~:~50, placed in an alumina crucible and then sealed in a quartz tube under vacuum (10$^{-3}$~mbar).  The whole assembly was placed in a box furnace and heated to 1000~$^\circ$C at a rate of 50~$^\circ$C/hr, left there for 10~hrs and then cooled to 500~$^\circ$C at a rate of 5~$^\circ$C/hr.  At this temperature the molten Pb flux was decanted.  Shiny plate-like crystals of typical size 1.5~mm$\times$1.5~mm$\times$0.1~mm were obtained.  The composition of the crystal was checked using energy dispersive x-ray (EDX) semiquantitative analysis using a JEOL scanning electron microscope (SEM).  The SEM scans were taken on a cleaved surface of a crystal.  The EDX gave the average elemental ratio Ba:Rh:As~=~20.4~:~40.8~:~38.8 which is consistent with an approximate 1:2:2 stoichiometry for the compound.  An SEM image of a typical as-grown crystal is shown in Fig.~\ref{Fig-Xtal}(a).  Laue back scattering measurements on the crystals showed that the largest surface of the plates was perpendicular to the $c$-axis. 
For crystal structure determination, single crystal x-ray diffraction measurements were done at temperature $T$~=~293~K using a Bruker CCD-1000 diffractometer with Mo $K_{\alpha}$ ($\lambda$~=~0.71073~\AA) radiation.  To look for structural phase transitions, single crystal x-ray diffraction measurements were performed on a standard four-circle diffractometer using Cu $K_\alpha$ radiation from a rotating anode x-ray source, selected by a Ge(1 1 1) monochromator.  For these measurements, a plate-like single crystal with dimensions of 2$\times$1.5$\times$0.2~mm$^3$ was selected.  The sample was mounted on a flat copper sample holder in a closed cycle displex cryogenic refrigerator with the (0 0 1)--(1 1 0) reciprocal lattice plane coincident with the scattering plane.  The diffraction patterns were recorded for temperatures between 10~K and 300~K and with a setup optimized for high resolution in transverse scans in the scattering plane.  The measured mosaicity of this crystal was 0.025$^\circ$ full width half maximum for both the (0 0 10) and (1 1 10) reflections at room temperature as shown in Fig.~\ref{Fig-Xtal}(b), indicating the good quality of the single crystal. 

The $\chi(T)$ was measured on a collection of aligned crystals with a total mass of 5.11~mg using a commercial Superconducting Quantum Interference Device (SQUID) magnetometer (MPMS5, Quantum Design), the standard four-probe $\rho(T)$ was measured with a current of amplitude $I$~=~2~mA at a frequency of 16~Hz, using the ac transport option of a commercial Physical Property Measurement System (PPMS5, Quantum Design).  The contacts were made with silver epoxy on a cleaved surface of a crystal.  The current $I$~=~2~mA was applied in the $ab$-plane.  The $C(T)$ was measured on a collection of three crystals of total mass 3.2~mg using the commercial PPMS.  

For the density of states (DOS) calculations, we have used the full potential linearized augmented plane wave (FP-LAPW) method with a local density approximation (LDA) functional.\cite{Perdew1992}  The difference in energy of 0.01~mRy/cell between successive iterations was used as a convergence criterion.  The value of $R_{\rm MT}$$K_{\rm max}$ [the smallest muffin tin (MT) radius multiplied by the maximum $k$ value in the expansion of plane waves in the basis set] which determines the accuracy of the basis set used, is set to 9.0.  The total Brillouin zone was sampled with 405 $k$-points in the irreducible Brillouin zone.  The employed MT radii are 2.4, 2.2 and 2.2 atomic units for Ba, Rh, and As, respectively.  The structural data (lattice constants and atomic positions) were taken from the previously reported single crystal structure of BaRh$_2$As$_2$.\cite{Hellmann2007}

\section{RESULTS}
\subsection{Single Crystal X-ray diffraction and Structure of BaRh$_2$As$_2$}
\label{sec:RES-structure}
%For X-ray diffraction measurements a few plate like crystals were glued onto a glass slide with vaseline.  The x-ray diffraction pattern so obtained is shown in Fig.~\ref{Fig-Xtal}(b).  Apart from a broad peak at low angles which is most likely the glass and some small reflections from a missalignment of the crystals, the only lines observed are the (00$l$) lines expected for the ThCr$_2$Si$_2$ type structure indicating that the square face of the crystals are perpendicular to the crystallographic $c$-axis.   
 
A small (0.35~mm$\times$0.35~mm$\times$0.07~mm) plate-like single crystal was used for crystal structure determination.  The initial cell parameters were taken as those previously reported for BaRh$_2$As$_2$ (ThCr$_2$Si$_2$ structure, $Z$~=~2 formula units/unit cell, space group $I4/mmm$).\cite{Hellmann2007}  The final cell parameters and atomic positions were calculated from a set of 861 strong reflections with good profiles in the range 2$\theta$~=~6$^\circ$--61$^\circ$.  The unit cell parameters were found to be $a$~=~$b$~=~4.0564(6)~\AA~ and $c$~=~12.797(4)~\AA.  These values are in good agreement with previously reported values for single crystalline BaRh$_2$As$_2$ [$a$~=~$b$~=~4.053(1) \AA, and $c$~=~12.770(3) \AA ].\cite{Hellmann2007}  There is only one atomic coordinate not constrained by symmetry requirements, the $z$ position for As.  Our value of $z$~=~0.3566(3) is also in good agreement with $z$~=~0.3569(1) reported previously by the full single crystal refinement.\cite{Hellmann2007}  The single crystals of BaRh$_2$As$_2$ have a high tendency to split/cleave into very thin plates perpendicular to the $c$-axis resulting in significant mosaicity/twinning.  This resulted in some reflections to be broad/extended. 
This and the unfavorable aspect ratio of the crystal for the appropriate absorption correction did not allow a full refinement of the structure with a reasonable $R$ factor.  

\begin{figure}[t]
\includegraphics[width=3in]{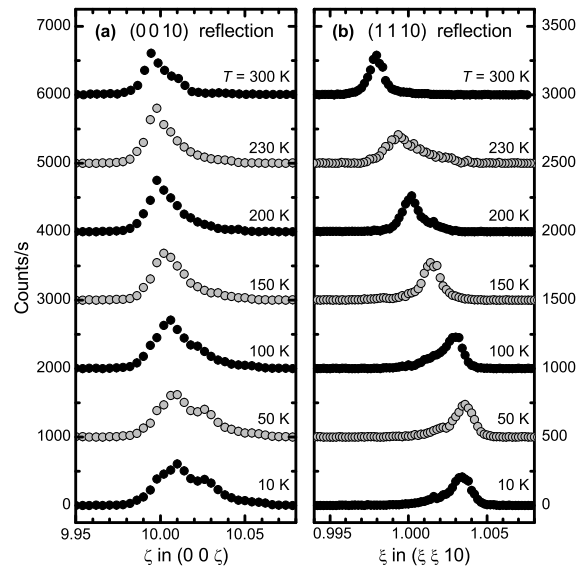}
\caption{(a) Longitudinal (0 0 $\zeta$) scans through the position of the (0 0 10) reflection for selected temperatures. The offset between each data set is 1000~counts/s. 
(b) Transverse ($\xi$ $\xi$ 0) scans through the position of the (1 1 10) reflection for selected temperatures. The offset between each data set is 500~counts/s.  In (a) and (b), the horizontal axes represent the scattering vectors relative to the alignment at $T$~=~200~K to illustrate the temperature dependence of the reflection positions.    
\label{Fig-LT-Xray}}
\end{figure}

Figures~\ref{Fig-LT-Xray}(a) and (b) show the results of our temperature dependent single crystal x-ray diffraction measurements which were done to check for the possibility of a structural phase transition in BaRh$_2$As$_2$.  Figure~\ref{Fig-LT-Xray}(a) shows the longitudinal (0 0 $\zeta$) scans through the (0 0 10) reflection at various temperatures between 10~K and 300~K relative to the alignment at $T$~=~200~K\@.  We did not observe any change in the shape of the (0 0 10) reflection as function of temperature.  The change in the position of the peak results mainly from lattice parameter changes between different temperatures.  The small peak at the shoulder of the (0 0 10) reflection at low temperatures [Fig.~\ref{Fig-LT-Xray}(a)] arises from a second grain with a slightly different orientation.  In contrast to the isostructural compounds $A$Fe$_2$As$_2$ ($A$~=~Ba, Sr, Ca),\cite{Ni2008, Yan2008,Ni2008a} transverse ($\xi$ $\xi$ 0) scans through the (1 1 10) reflections [Fig.~\ref{Fig-LT-Xray}(b)] show no splitting and/or abrupt changes in reflection position over the temperature range between 10 and 300 K, indicating the absence of the structural transition reported for the isostructural $A$Fe$_2$As$_2$ ($A$~=~Ba, Sr, Ca) compounds.\cite{Ni2008, Yan2008,Ni2008a}  This is consistent with the macroscopic measurements discussed later.

\subsection{Density Of States}
\label{DOS}
\begin{figure}[t]
\includegraphics[width=3.in]{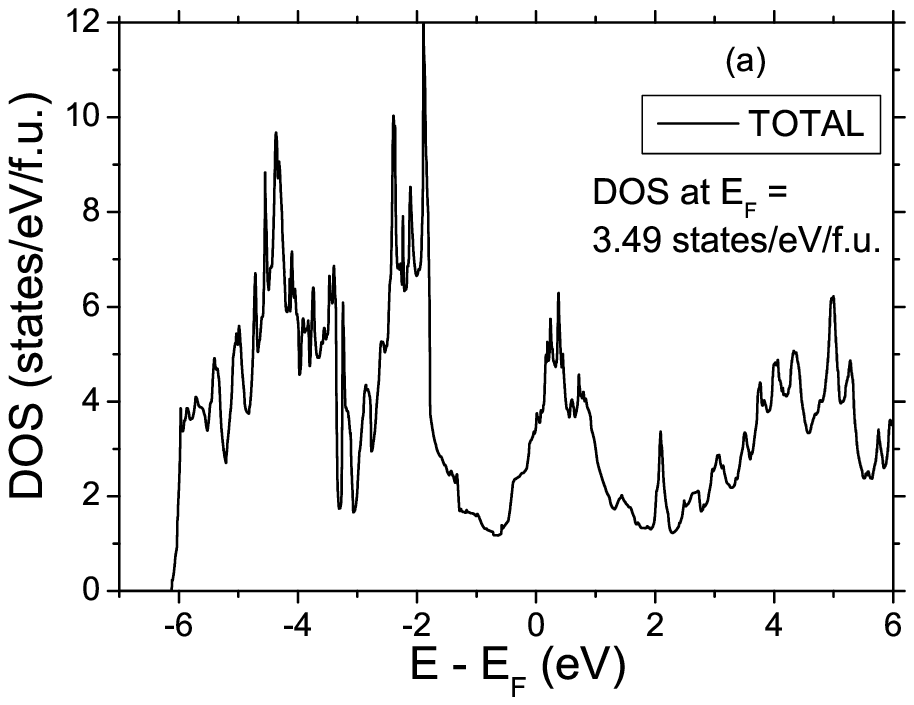}
\includegraphics[width=3.in]{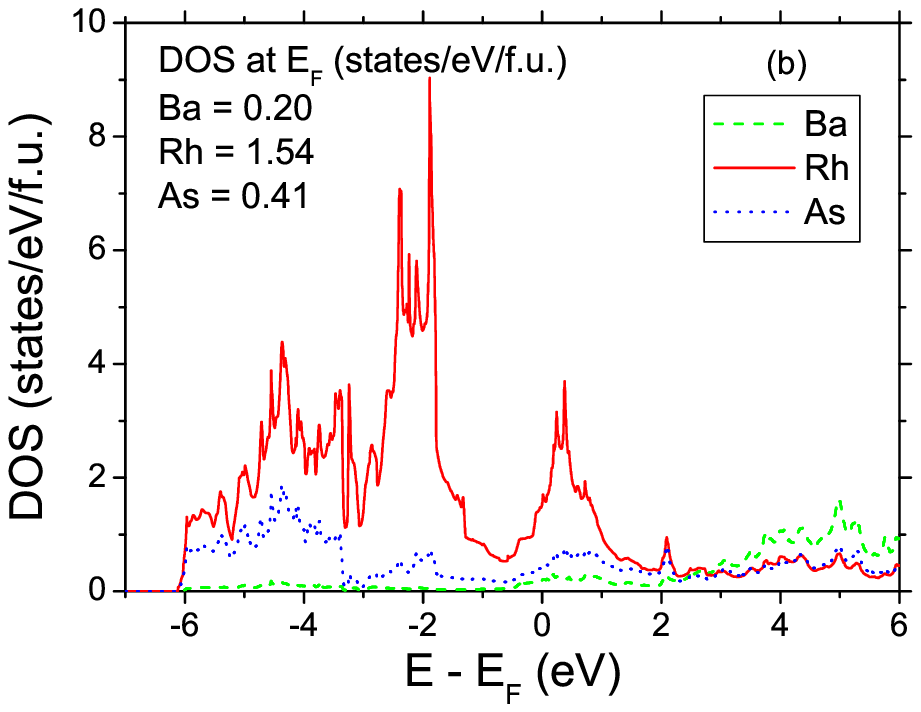}
\caption{ (a) The total density of states for both spin directions for BaRh$_2$As$_2$.  (b) Partial density of states of Ba, Rh, and As atoms in BaRh$_2$As$_2$.
\label{FigDOS}}
\end{figure}

\begin{figure}[t]
\includegraphics[width=3.in]{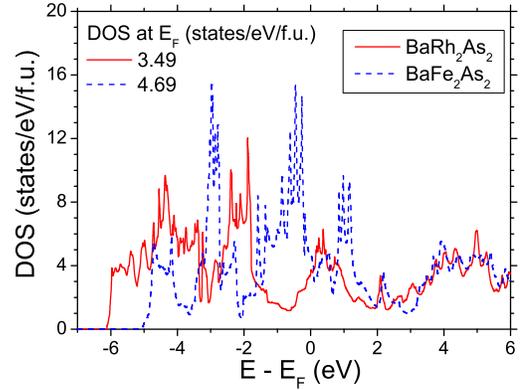}
\caption{Comparison of the total density of states for both spin directions for BaRh$_2$As$_2$ (solid curve) with those for BaFe$_2$As$_2$ (dashed curve).  
\label{Fig-DOS-total-RhvsFe}}
\end{figure}

The calculated total density of states (DOS) $N(E_{\rm F})$ for BaRh$_2$As$_2$ for both spin directions versus energy $E$ measured relative to the Fermi energy $E_{\rm F}$, is shown in Fig.~\ref{FigDOS}(a).  We calculate $N(E_{\rm F})$~=~3.49~states/eV~f.u. (f.u.~=~formula unit) for both spin directions for BaRh$_2$As$_2$.  Figure~\ref{FigDOS}(b) shows the partial DOS for the individual atoms.  The maximum contribution to the total $N(E_{\rm F})$ is from the Rh 4$d$ states with small but significant contributions from Ba and As.  Since rather small MT radii were employed, wave functions for the 4$d$ Rh states extend out into the interstitial region.  This results in a significant DOS in the interstitial region which is not accounted for in the partial DOS estimates.  This interstitial DOS is however accounted for in calculation of the total DOS.  Therefore, the partial DOS shown in Fig.~\ref{FigDOS}(b) do not add up to the total DOS shown in Fig.~\ref{FigDOS}(a).  It is of interest to compare the DOS versus energy for BaRh$_2$As$_2$ with that of BaFe$_2$As$_2$.  Figure~\ref{Fig-DOS-total-RhvsFe} shows the total DOS for both spin directions for BaRh$_2$As$_2$ (solid curve) and BaFe$_2$As$_2$ (dashed curve).  The band structures for the two materials are seen to be qualitatively quite similar.  It appears that the bands for BaRh$_2$As$_2$ are shifted down (and stretched) in energy relative to the bands for BaFe$_2$As$_2$.  For BaFe$_2$As$_2$, if the Fermi energy $E_{\rm F}$ moves down by about 0.5~eV on hole doping, there is a large increase in the DOS (almost 4--5 times) compared to the parent compound.  To move the $E_{\rm F}$ for the BaRh$_2$As$_2$ material to the corresponding band located at about $-$2~eV below $E_{\rm F}$ would evidently require a much larger amount of doping.


\subsection{Magnetic Susceptibility}
\label{sec:RES-susceptibility}
\begin{figure}[t]
\includegraphics[width=3in]{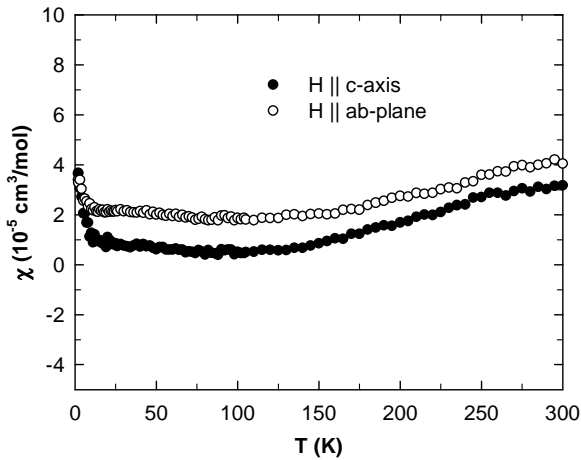}
\caption{The magnetic susceptibility $\chi$ versus temperature $T$ for BaRh$_2$As$_2$ for the magnetic field $H$ applied parallel and perpendicular to the tetragonal $c$-axis.  
\label{Figsus}}
\end{figure}

The magnetic susceptibilities for BaRh$_2$As$_2$ measured with an applied magnetic field $H$~=~2~T parallel to the $c$-axis $\chi_{\rm c}(T)$ and with $H$ parallel to the $ab$-plane $\chi_{\rm ab}(T)$ are shown in Fig.~\ref{Figsus} (after correcting for contributions from the sample holder).  The measurements were made on a collection of four $c$-axis aligned crystals of total mass 5.11~mg.  Both $\chi_{\rm c}(T)$ and $\chi_{\rm ab}(T)$ are almost temperature independent and have very small average values $\chi_{\rm c}(T)$~=~1.5$\times 10^{-5}$~cm$^3$/mol and $\chi_{\rm ab}(T)$~=~3$\times 10^{-5}$~cm$^3$/mol. There is a small anisotropy in the whole temperature range with $\chi_{\rm ab}$/$\chi_{\rm c}\approx2$.  The data are somewhat noisy, most likely due to both the small absolute values of the susceptibility and the small crystal mass used for the measurements.  The powder averaged susceptibility $\chi(T)$ can be calculated for a tetragonal system as
\begin{equation}
\chi~=~{2\chi_{\rm ab}+\chi_{\rm c}\over 3}~.
\label{Eq-powderaverage}
\end{equation}
\noindent
From the average values of $\chi_{\rm ab}$ and $\chi_{\rm c}$, we estimate an average value $\chi$~=~2.5$\times 10^{-5}$~cm$^3$/mol from 10~K to 300~K\@.  

The susceptibility $\chi(T)$ can be written as 

\begin{equation}
\chi~=~\chi_{\rm core}+\chi_{\rm L}+\chi_{\rm VV}+\chi_{\rm P}~,
\label{Eqcontributionstochi}
\end{equation}
\noindent
where $\chi_{\rm core}$ is the diamagnetic orbital contribution from the localized core electrons (ionic or atomic), $\chi_{\rm L}$ is the Landau orbital diamagnetism of the conduction electrons, $\chi_{\rm VV}$ is the Van Vleck paramagnetic orbital contribution and $\chi_{\rm P}$ is the Pauli paramagnetic spin susceptibility of the conduction electrons.  We evaluated $\chi_{\rm P}$ and $\chi_{\rm VV}$ from electronic structure calculations using\cite{Oh1976} 
\begin{equation}
\chi_{\rm P}~=~\mu_{\rm B}^2 N(E_{\rm F})~, 
\label{EqDOSCHIP1}
\end{equation}
\noindent
where $\mu_{\rm B}$ is the Bohr magneton and $N(E_{\rm F})$ is the bare density of states at the Fermi level for both spin directions, and 
\begin{equation}
\chi_{\rm VV}~=~\mu_{\rm B}^2 \sum_{n,n',\vec{k}}{f_{n\vec{k}} - (1-f_{n'\vec{k}})\over E_{n\vec{k}} - E_{n'\vec{k}}}~ |<n\vec{k}|L_z|n'\vec{k}>|~, 
\label{EqDOSCHIVV}
\end{equation}
\noindent
where $f$ is the Fermi-Dirac distribution function and $L_z$ is the $z$ component of the angular momentum operator.\cite{Oh1976}  We have calculated the anisotropic $\chi_{\rm VV}$ for magnetic field $H$ in the (001) and (110) directions.  

The diamagnetic contribution $\chi_{\rm D}$~=~$\chi_{\rm core}+\chi_{\rm L}$ to the total susceptibility was also estimated.  Since diamagnetic susceptibility is problematic to calculate in extended systems because of inter-cell currents, we take the value from calculations of the neutral atoms and ignore the solid state effects.  This gives a lower limit estimate of $\chi_{\rm D}$.  It also misses any anisotropy arising from inter-cell currents.

The results of these calculations are summarized in Table~\ref{tabchi}.
\begin{table}

\caption{\label{tabchi}
Estimated contributions to the total susceptibility $\chi$ in units of $10^{-6}$~cm$^3$/mole}
\begin{ruledtabular}
\begin{tabular}{|l|cc|}
contribution&Value&\\ \hline 
$\chi_{\rm P}$&112& \\\hline  
$\chi_{\rm D}$&$-$222& \\ \hline
$\chi_{\rm VV}$&$H$ $\|$ (001)& $H$ $\|$ (110) \\
&153& 138 \\ \hline
$\chi$&43&28\\
\end{tabular}
\end{ruledtabular}
\end{table}
It can be seen that although the individual contributions $\chi_{\rm P}$, $\chi_{\rm D}$, and $\chi_{\rm VV}$ are large there is a large cancellation between the positive $\chi_{\rm P}$ and $\chi_{\rm VV}$, and the diamagnetic $\chi_{\rm D}$ which results in a small net $\chi$.  Although the total $\chi$ from the calculations is of the same order as the experimentally measured value, the calculations could not quantitatively reproduce the experimental numbers and the anisotropy (Fig.~\ref{Figsus}).  Since these calculations did not take into account the anisotropy of $\chi_{\rm D}$ associated with valence electrons, the above comparison strongly suggests that $\chi_{\rm D}$ also has a large anisotropy, with a larger value for fields along the $c$-axis. 

\subsection{Resistivity}
\label{sec:RES-resistivity}
\begin{figure}[t]
\includegraphics[width=3in]{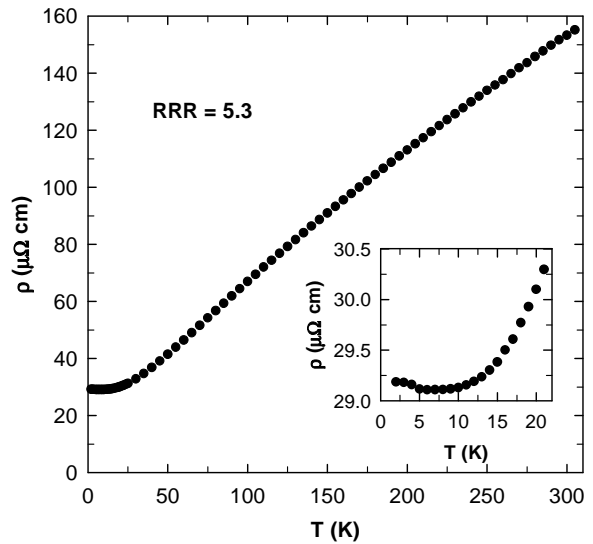}
\caption{The resistivity $\rho$ versus temperature $T$ in the $ab$-plane for a crystal of BaRh$_2$As$_2$ between 2~K and 300~K\@.  The inset shows the $\rho(T)$ data between 2~K and 22~K on an expanded scale to highlight the residual resistivity $\rho_0$ and the slight upturn below about 5~K\@.  As noted, the residual resistivity ratio is RRR~$\equiv$~$\rho(310~{\rm K})$/$\rho(2~{\rm K})$~=~5.3.
\label{Figres}}
\end{figure}

The electrical resistivity $\rho$ versus temperature $T$ in the $ab$-plane is shown in Fig.~\ref{Figres}.  The $\rho(T)$ data show typical metallic behavior with a monotonic decrease with decreasing $T$ between 310~K and 10~K\@.  Below $T$~=~10~K $\rho(T)$ saturates to a residual resistivity value $\rho(2~{\rm K})$~=~29(3)$\mu \Omega$~cm.  There is however a slight upturn in $\rho(T)$ at the lowest temperature as seen in the inset in Fig.~\ref{Figres}.  A corresponding upturn is also observed at low temperatures in the heat capacity $C(T)$ data discussed below.  This upturn in both the $\rho(T)$ and $C(T)$ data at low temperatures may be due to the presence of a small amount of paramagnetic impurity in the material.  However, it cannot be ruled out that this behavior is intrinsic to the material.  The residual resistivity ratio RRR~=~$\rho(310~{\rm K})$/$\rho(2~{\rm K})$~=~5.3 for our BaRh$_2$As$_2$ crystal is typical of what is observed for single crystals of the isostructural $A$Fe$_2$As$_2$ ($A$~=~Ba, Sr, Ca, and Eu) materials.\cite{Krellner2008,Ni2008,Yan2008,Ronning2008,Jeevan2008,Wang2008,Luo2008}

\subsection{Hall effect}
\label{sec:Hall}
\begin{figure}[t]
\includegraphics[width=3in]{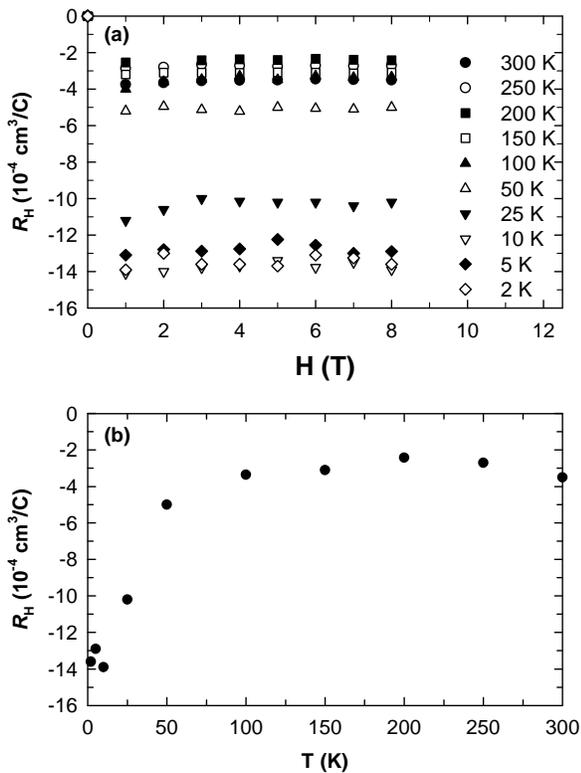}
\caption{(a) The Hall coefficient $R_{\rm H}$ versus magnetic field $H$ measured at various temperatures.  (b) The Hall coefficient $R_{\rm H}$ versus temperature $T$ in a magnetic field of 8~T\@.  
\label{FigHall}}
\end{figure}

Figure~\ref{FigHall}(a) shows the dependence of the Hall coefficient $R_{\rm H}$ on magnetic field $H$ at various temperatures.  The $R_{\rm H}$ is negative at all fields and temperatures which indicates that electron conduction dominates the electronic transport.  The $R_{\rm H}$ is approximately constant with magnetic field at all temperatures.  From $R_{\rm H}$ one obtains a lower limit estimate of the carrier density using the single-band expression \cite{hurd1972}
$$R_{\rm H} = {1\over nq}~,$$
where $n$ is the carrier density and $q$ is the charge of the current carriers.  Using the value $R_{\rm H}$~=~3.5$\times$10$^{-4}$~cm$^3$C$^{-1}$ at 300~K and 8~T and $q$~=~$-e$ where $e$ is the magnitude of the charge of the electron, we get $n$~=~1.8$\times$10$^{22}$~cm$^{-3}$\@.  One can get an estimate of the density of states at the Fermi energy $N(E_{\rm F})$ using the single-band relation \cite{Kittel}    
\begin{equation}
N(E_{\rm F}) = {m V_{\rm f. u.} (3n)^{1\over 3}\over \hbar^2\pi^{4\over 3}}~,
\end{equation}
where $m$ is the free electron mass, $V_{\rm f.u.}$~=~210.5~\AA$^3$ is the volume per formula unit, and $\hbar$ is Planck's constant divided by 2$\pi$\@.  With the value of $n$ obtained above one gets $N(E_{\rm F})$~=~2.3~states/(eV~f.u.) for both spin directions which is smaller than the 3.49~states/(eV~f.u.) obtained above from band structure calculations.  This suggests that both electrons and holes contribute to the electrical conduction and the net carrier density estimated from the Hall measurements is smaller than the actual value for the material, and/or that the conduction electron effective mass is greater than the free electron mass.

Figure~\ref{FigHall}(b) shows the plot of $R_{\rm H}$ versus temperature $T$ at a magnetic field of 8~T\@.  The $R_{\rm H}$ is almost constant between T~=~50~K and 300~K\@.  Below $T$~=~50~K however, it decreases strongly by a factor of 4--5.  In a single band model the $R_{\rm H}$ is expected to be temperature independent if the scattering rate $\tau$ is isotropic.  However, in a two-band model, the $R_{\rm H}$ could be temperature dependent if, for example, the scattering rates of the two bands have a different temperature dependence or the fractions of charge carriers in the two bands change with temperature.\cite{hurd1972}  

\subsection{Heat Capacity}
\label{sec:RES-heatcapacity}
\begin{figure}[t]
\includegraphics[width=3in]{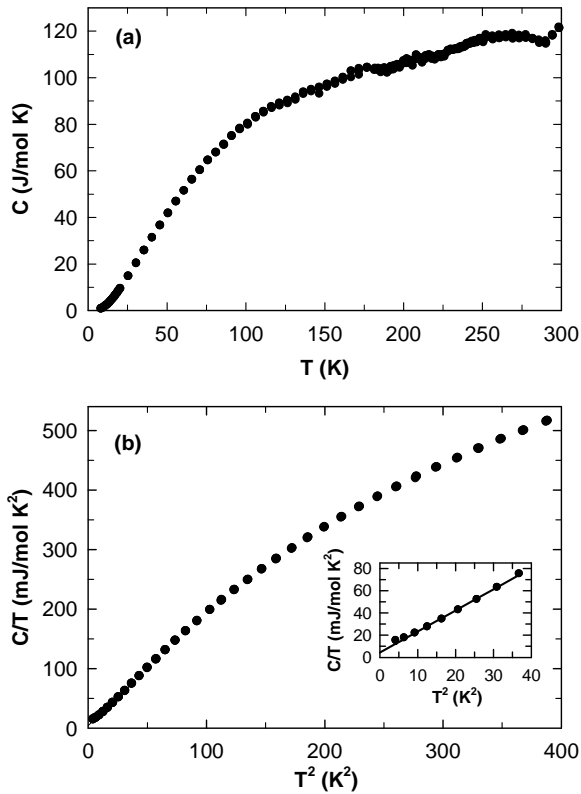}
\caption{(a) Heat capacity $C$ versus temperature $T$ for BaRh$_2$As$_2$ crystals between 2~K and 300~K\@.  (b) $C/T$ versus $T^2$ data between 2~K and 20~K\@.  The inset shows the $C(T)/T$ versus $T^2$ data below $T$~=~6~K\@.  The solid line through the data in the inset is a fit by the expression $C/T$~=~$\gamma$+$\beta T^2$.  
\label{FigHC}}
\end{figure}

The heat capacity $C$ versus temperature $T$ of BaRh$_2$As$_2$ between 1.8~K and 300~K is shown for a collection of three crystals in Fig.~\ref{FigHC}(a).  The data at high temperatures are somewhat noisy because of the small mass $m$~=~3.2~mg of the crystals used for the measurement.  However, the heat capacity at room temperature $C(300~{\rm K})\approx$~120~J/mol~K is close to the classical Dulong Petit value $C$~=~125~J/mol~K expected for BaRh$_2$As$_2$.  Below $T$~=~100~K the quality of the data is better.  There is no signature of any phase transition in the temperature range of our measurements.  

Figure~\ref{FigHC}(b) shows the $C(T)/T$ versus $T^2$ data between 2~K and 20~K\@.  The $C(T)/T$ data show curvature in the whole temperature range shown.  This suggests that there may be anharmonic contributions to the lattice heat capacity even at low temperatures.  Only below about 10~K do the $C(T)/T$ data show a quasi-linear dependence on $T^2$.  At the lowest temperature the $C/T$ shows a tendency of saturating to a value $C(2~{\rm K})/T$~=~15.5~mJ/mol~K$^2$ as shown in the inset in Fig.~\ref{FigHC}(b) where the $C/T$ versus $T$ data are shown below $T$~=~6~K\@.  From fits to the $C/T$ versus $T$ data up to 10~K by the expression $C/T$~=~$\gamma$+$\beta T^2$, where $\gamma$ is the Sommerfeld coefficient of the electronic heat capacity and $\beta$ is the coefficient of the lattice heat capacity, we obtain $\gamma$~=~4.7(9)~mJ/mol~K$^2$ and $\beta$~=~1.93(4)~mJ/mol~K$^4$.     

The density of states at the Fermi energy for both spin directions $N(E_{\rm F})$ can be estimated from the above value of $\gamma$ using the relation \cite{Kittel}
\begin{equation}
\gamma~=~{\pi^2 \over 3}k_{\rm B}^2N(E_{\rm F})(1+\lambda_{\rm ep}) ~.
\label{EqDOSHC}
\end{equation}
\noindent
where $\lambda_{\rm ep}$ is the electron-phonon coupling constant.  Setting $\lambda_{\rm ep}$~=~0 as a first approximation, one gets $N(E_{\rm F})$ ~=~2.0(4)~states/(eV~f.u.).  This is a smaller than the value predicted from band structure calculations $N(E_{\rm F})$~=~3.49~states/(eV~f.u.). 

From the value of $\beta$~=~1.93(4)~mJ/mol~K$^4$ estimated above one can obtain the Debye temperature $\Theta_{\rm D}$ using the expression \cite{Kittel}
\begin{equation}
\Theta_{\rm D}~=~\bigg({12\pi^4Rn \over 5\beta}\bigg)^{1/3}~, 
\label{EqDebyetemp}
\end{equation}
\noindent
where $R$ is the molar gas constant and $n$ is the number of atoms per formula unit (\emph{n}~=~5 for BaRh$_2$As$_2$).  We obtain $\Theta_{\rm D}$~=~171(2)~K for BaRh$_2$As$_2$.  The values of $\gamma$ and $\Theta_{\rm D}$ estimated for BaRh$_2$As$_2$ are similar to what has been reported for the isostructural compounds $A$Fe$_2$As$_2$ ($A$~=~Ca, Sr, Ba, and Eu).\cite{Ni2008a,Yan2008,Rotter2008,Ni2008,Jeevan2008}

\noindent

\section{CONCLUSION}
\label{sec:CON}
We have synthesized single crystalline samples of the layered rhodium arsenide BaRh$_2$As$_2$ and characterized them using single crystal x-ray diffraction, anisotropic magnetic susceptibility versus temperature, resistivity versus $T$, and heat capacity versus $T$ measurements between 2~K and 300~K\@.  The single crystal structure determination confirms that BaRh$_2$As$_2$ crystallizes in the tertagonal ThCr$_2$Si$_2$ type structure with lattice parameters $a$~=~$b$~=~4.0564(6)~\AA~ and $c$~=~12.797(4)~\AA.  Single crystal x-ray diffraction meaasurements down to 10~K did not show any evidence for a structural transition.  The density of states calculations give a total density of states for both spin directions $N(E_{\rm F})$~=~3.49~states/(eV~f.u.) with maximum contribution from the Rh 4$d$ states.  The $\chi(T)$ is small and temperature dependent.   The small $\chi(T)$ suggests a small density of states.  This is supported by the $C(T)$ data which give a small Sommerfeld coefficient $\gamma$~=~4.7(9)~mJ/mol~K$^2$.

\begin{acknowledgments}
We thank P. C. Canfield for assistance with decanting the flux from the crystals.  Work at the Ames Laboratory was supported by the Department of Energy-Basic Energy Sciences under Contract No.\ DE-AC02-07CH11358.  
\end{acknowledgments}

\end{document}